\documentclass[twocolumn,aps,prd,nofootinbib,natbib]{revtex4-2}

\usepackage[utf8]{inputenc}
\usepackage[T1]{fontenc}
\usepackage{graphicx}
\usepackage{amssymb}
\usepackage{amsmath}
\usepackage{xcolor}
\usepackage{hyperref}
\usepackage{textcase}

\hypersetup{
    colorlinks=true,
    linkcolor=blue,
    filecolor=magenta,      
    urlcolor=blue,
    citecolor=blue,
}
\begin{document}

\title{\MakeTextUppercase{Wormhole Dynamics: Nonlinear Collapse and Gravitational-Wave Emission
}}

\author{\MakeTextUppercase{Nikita M. Shirokov}}
\email{shirokov.nm@phystech.edu}
\affiliation{\textit{Independent Researcher, Moscow, Russia}}
\date{\today}

\begin{abstract}
We present 3D numerical-relativity evolutions of the unstable Ellis--Bronnikov wormhole using \texttt{GRTeclyn}, starting from exact isotropic initial data for the coupled Einstein--phantom-scalar system. With a flat initial lapse ($\alpha=1$) and full phantom support, truncation-level noise eventually drives the rarefactive instability and rapid throat expansion. To force a clean collapse while breaking spherical symmetry, we reduce the phantom stress-energy support to $S_{\rm support}=0.5$ and add a quadrupolar scalar-field perturbation ($A_\phi=+0.02$, $\sigma_\phi=0.5$). The resulting compressive evolution forms a trapped surface and emits a gravitational-wave signal whose peak propagates between extraction radii at $v\approx c$, distinct from superluminal CCZ4 constraint modes. After horizon formation the swallowed phantom matter triggers a violent rebound (``phantom bounce'') that launches an outward curvature shock. For the moderate perturbation amplitude simulated here, an intermediate-mass ($10^3\,M_\odot$) wormhole at $D=1$~Mpc falls slightly below the Advanced LIGO design sensitivity; detection requires either closer sources, larger initial asymmetries, or next-generation detectors.
\end{abstract}

\maketitle

\section{Introduction}
Wormholes---topological bridges connecting distinct spacetime regions---have been studied since Flamm~\cite{flamm16}, Einstein and Rosen~\cite{einstein35}, and Wheeler~\cite{misner57}. The canonical traversable solution, supported by a phantom scalar field, was found independently by Ellis~\cite{ellis73} and Bronnikov~\cite{bronnikov73}.

\begin{figure}[htp]
\centering
\includegraphics[width=\linewidth]{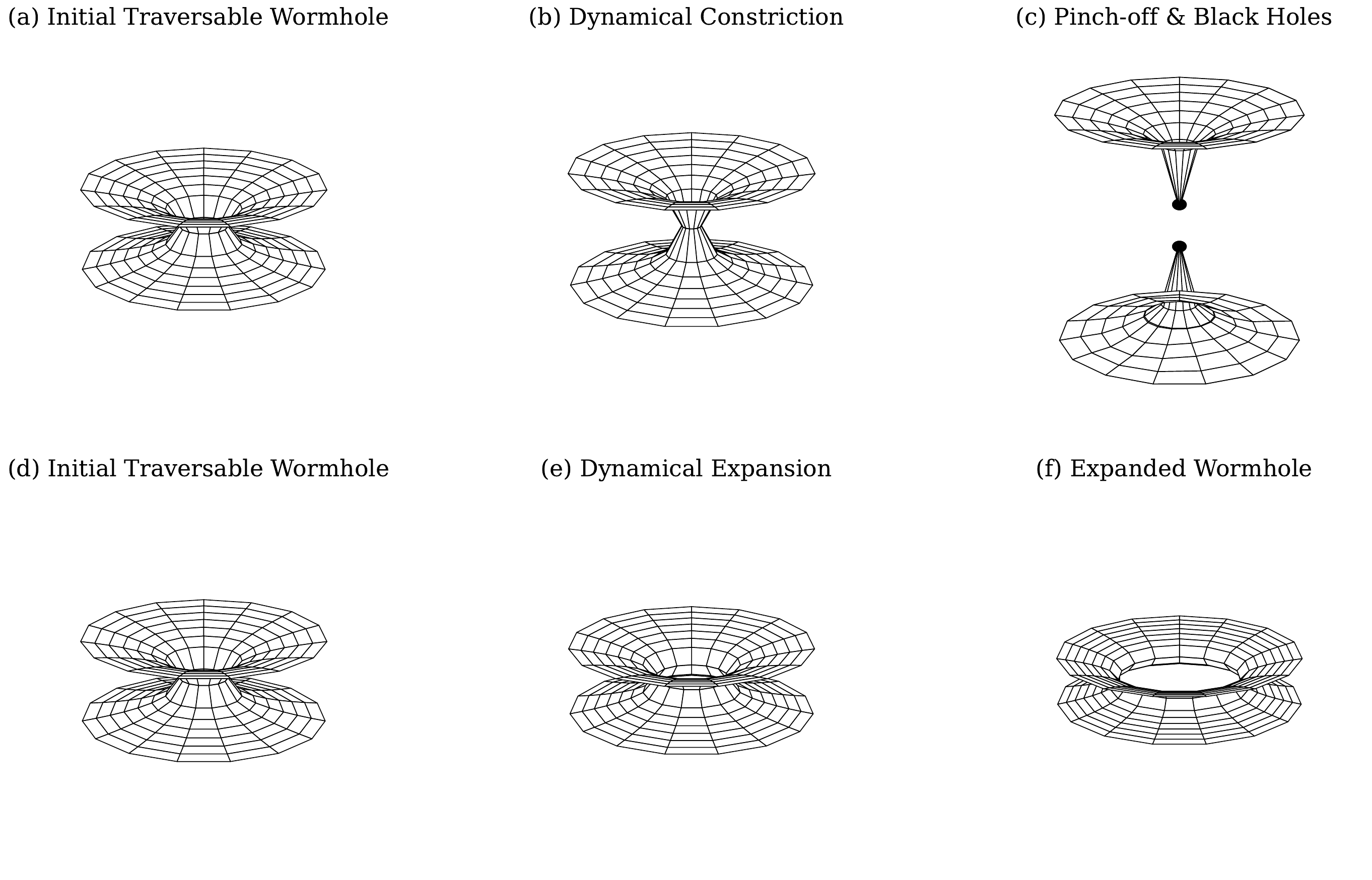}
\caption{Embedding diagram illustrating the two instability pathways: compressive collapse (top) and rarefactive expansion (bottom). Note that classical GR evolution mathematically preserves the underlying spatial topology ($\mathbb{R} \times S^2$). The ``pinch-off'' depicts a metric degeneration where the throat's areal radius shrinks and an apparent horizon forms, severing causal contact between the two asymptotic regions without tearing the manifold itself.}
\label{fig:wormhole_collapse_stages}
\end{figure}

Static traversable wormholes require exotic matter violating the null energy condition~\cite{morris88}. If such defects were produced in spacetime foam and stretched to macroscopic scales during inflation~\cite{kardashev07,garcia16}, the relevant question is their nonlinear dynamical fate. Shinkai and Hayward~\cite{shinkai02} showed in 1D spherical symmetry that the Ellis--Bronnikov geometry is highly unstable: rarefactive perturbations trigger expansion, while compressive ones force collapse (Fig.~\ref{fig:wormhole_collapse_stages}). However, spherical symmetry prohibits gravitational radiation (the lowest radiative multipole is $\ell=2$), so 3D simulations are needed to quantify any gravitational-wave signal.

We evolve the coupled Einstein--phantom-scalar system in 3D from exact Ellis--Bronnikov initial data and extract the gravitational-wave signal via the Weyl scalar \(\Psi_4\) at multiple radii. A propagation-speed test separates physical gravitational radiation (\(v\approx c\)) from superluminal CCZ4 constraint-damping features, and the contrast between unperturbed and perturbed runs provides a built-in null test.

Following recent 3D studies of exotic geometries~\cite{clough24}, we adopt an approach in which the exotic source is reduced at $t{=}0$, initiating a dynamical relaxation problem with an initial Hamiltonian constraint defect. Furthermore, any primordial wormhole that survived inflation would also acquire aspherical volume changes from tidal interactions, reheating, and cosmological expansion; the quadrupolar scalar-field profile perturbation we impose at $t{=}0$ models this physically expected post-inflationary state.

With gravitational-wave astronomy now operational~\citep{abbott16,abbott18,abbott23}, understanding the dynamics of Exotic Compact Object candidates~\cite{cardoso19} is essential for any detection strategy.

\section{Formalism and Methodology}

\subsection{3+1 Decomposition and the CCZ4 Formulation}
We follow the standard numerical-relativity approach to solving the Einstein Equations,
\begin{equation}
G_{\mu\nu} = R_{\mu\nu} - \frac{1}{2} R g_{\mu\nu} = 8\pi T_{\mu\nu},
\end{equation}
as an initial value problem, using the 3+1 decomposition~\cite{baumgarte10} and setting $G = c = 1$. The four-dimensional spacetime manifold, which topologically is $\mathcal{M} = \mathbb{R}^2 \times S^2$, is foliated into a family of spacelike hypersurfaces $\Sigma_t$ with topology $\mathbb{R} \times S^2$, parametrized by a time coordinate $t$. The line element is written in terms of the lapse function $\alpha$, the shift vector $\beta^i$ (which controls how the spatial coordinates propagate between time slices), and the spatial metric $\gamma_{ij}$ as:
\begin{equation}
    ds^2 = \left(-\alpha^2 + \beta_i \beta^i \right) dt^2 + 2 \beta_i dt dx^i + \gamma_{ij} dx^i dx^j.
\end{equation}
The dynamical evolution is governed by the coupled Einstein--scalar-field system, implemented within the \texttt{GRTeclyn} framework using the conformal and covariant Z4 (CCZ4) formulation with matter source terms. Our objective is to start from the exact Ellis--Bronnikov wormhole solution---geometry \emph{and} its phantom scalar field support---and evolve the full coupled system self-consistently, so that the initial data satisfy the Hamiltonian and momentum constraints to truncation error.

\subsection{Initial Data: Ellis--Bronnikov Wormhole in Isotropic Coordinates}

The general static, spherically symmetric traversable wormhole is described by the Morris--Thorne line element~\cite{morris88}:
\begin{equation}
ds^2 = -e^{2\Phi(r)} dt^2 + \frac{dr^2}{1 - b(r)/r} + r^2 (d\theta^2 + \sin^2\theta d\phi^2).
\end{equation}
Setting the redshift function to zero ($\Phi(r) = 0$) and choosing the shape function $b(r) = b_0^2/r$ yields the Ellis--Bronnikov wormhole~\cite{ellis73, bronnikov73}, a canonical massless phantom-scalar-field solution. As emphasized in Ellis's original formulation, the vanishing redshift ensures the initial unperturbed geometry is a geodesically complete, horizonless manifold. This property sets the proper time of Eulerian observers equal to coordinate time at $t=0$, and eliminates radial tidal forces at the throat; $b_0$ is the throat radius; and the choice of $b(r)$ guarantees asymptotic flatness as $r \to \infty$.

In standard Schwarzschild-like coordinates, the metric component $g_{rr} = (1 - b_0^2/r^2)^{-1}$ diverges at the throat ($r=b_0$), and this chart only covers one asymptotic half of the manifold. To resolve this coordinate singularity and map the entire $\mathbb{R} \times S^2$ spatial manifold onto a single punctured Euclidean grid ($\mathbb{R}^3 \setminus \{0\}$), we transform to isotropic coordinates. We introduce the isotropic radius $\bar{r}$, related to the proper distance $\ell = \pm \sqrt{r^2 - b_0^2}$ via the relation $\ell = b_0 \sinh(\ln(2\bar{r}/b_0))$. 

To initialize the wormhole on a 3D Cartesian grid, we apply this isotropic coordinate transformation. Expanding the hyperbolic relation and using $r^2 = \ell^2 + b_0^2$ yields the explicit radial coordinate map $r = \bar{r} \left( 1 + \frac{b_0^2}{4\bar{r}^2} \right)$. Substituting this into the standard line element yields a conformally flat representation of the spatial geometry of the Ellis--Bronnikov wormhole~\cite{nandi2008energetics}:
\begin{equation}
    ds^2 = -dt^2 + \left( 1 + \frac{b_0^2}{4\bar{r}^2} \right)^2 \big[ d\bar{r}^2 + \bar{r}^2 (d\theta^2 + \sin^2\theta d\phi^2) \big].
\end{equation}
From this isotropic form, the physical spatial metric $\gamma_{ij}$ can be explicitly extracted. Comparing this to the standard 3+1 conformal decomposition $\gamma_{ij} = \psi^4 \tilde{\gamma}_{ij}$ over a flat background ($\tilde{\gamma}_{ij} = \delta_{ij}$), the CCZ4 conformal factor $\psi$ is strictly given by:
\begin{equation}
    \psi = \sqrt{1 + \frac{b_0^2}{4\bar{r}^2}}.
\end{equation}
Unlike the Brill--Lindquist black-hole puncture~\cite{brill63}, where $\psi$ diverges at a point singularity, the conformal factor here represents a genuine topological bridge. In this chart, the two asymptotically flat ends are treated differently to fit on the grid: the primary asymptotic region (``primary universe'') extends to $\bar{r} \to \infty$ ($\psi \to 1$), the throat is a 2-sphere located at $\bar{r} = b_0/2$, and the secondary asymptotic region (``secondary universe'') is compactified such that its spatial infinity maps exactly to the grid origin $\bar{r} \to 0$ ($\psi \to \infty$). Here, the diverging conformal factor geometrically restores the infinite proper volume of the secondary region on the finite Cartesian grid.

\subsection{Transformation to CCZ4 Variables and Regularization}

Directly evolving the physical metric $\gamma_{ij}$ or the standard conformal factor $\psi$ would immediately crash a numerical code due to the $\bar{r} \to 0$ coordinate divergence. The CCZ4 formulation elegantly bypasses this by factoring out the divergent volume element. The physical spatial metric is decomposed into a well-behaved conformal metric $\tilde{\gamma}_{ij}$ with unit determinant, and a regularized conformal variable $\chi$:
\begin{align}
    \tilde{\gamma}_{ij} &= \psi^{-4} \gamma_{ij} = \delta_{ij}, \\
    \chi &= \psi^{-4} = \left( 1 + \frac{b_0^2}{4\bar{r}^2} \right)^{-2}.
\end{align}
While $\psi \to \infty$ at the compactified origin, its inverse power $\chi$ remains perfectly finite. To strictly avoid division-by-zero floating-point errors on the discrete computational grid near $\bar{r}=0$, the exact initial data for $\chi$ is algebraically rewritten and implemented as:
\begin{equation}
    \chi = \left( \frac{4\bar{r}^2}{4\bar{r}^2 + b_0^2} \right)^{2}.
\end{equation}
This formulation ensures that $\chi$ smoothly vanishes at the origin ($\chi \to 0$). By evolving $\chi$ rather than $\gamma_{ij}$, the CCZ4 framework adopts the ``moving-puncture'' philosophy, providing a completely regular, numerically stable way to dynamically evolve the infinite secondary universe as a single point on the Cartesian grid.

\subsection{Phantom Scalar Field and Self-Consistent Matter Evolution}

The Ellis--Bronnikov wormhole is supported by a massless phantom (ghost) scalar field $\phi$. As originally demonstrated by Ellis~\cite{ellis73}, reversing the coupling polarity between this field and the spacetime geometry produces a strictly negative-definite energy density, providing the necessary anti-gravity to keep the throat open. The resulting stress-energy tensor used for dynamical evolutions~\cite{shinkai02, gonzalez09} is
\begin{equation}
    T_{\mu\nu} = -\left(\nabla_\mu\phi\,\nabla_\nu\phi - \frac{1}{2}g_{\mu\nu}\nabla^\alpha\phi\,\nabla_\alpha\phi - g_{\mu\nu}\,V(\phi)\right),
\end{equation}
where the overall minus sign relative to the canonical scalar field ensures violation of the null energy condition, as required for a traversable wormhole. By enforcing the contracted Bianchi identity $\nabla_\mu T^{\mu\nu}=0$, the scalar field obeys the standard Klein--Gordon equation:
\begin{equation}
    \Box\phi = \frac{dV}{d\phi},
\end{equation}
which is identical in form to that of a normal scalar field; only the gravitational coupling (the sign of $T_{\mu\nu}$) is reversed.

In the 3+1 decomposition, the scalar field is represented by the field value $\phi$ and its conjugate momentum $\Pi \equiv n^\mu\nabla_\mu\phi$, where $n^\mu$ is the unit timelike normal. This sign convention matches the evolved variable \texttt{Pi} in \texttt{GRTeclyn} (note that some references define $\Pi=-n^\mu\nabla_\mu\phi$). Both are evolved alongside the CCZ4 geometric variables at every Runge--Kutta substep via the \texttt{CCZ4RHSWithMatter} module in \texttt{GRTeclyn}, ensuring full self-consistency between the matter and gravitational sectors.

For the exact Ellis--Bronnikov solution with $V(\phi)=0$, the scalar field profile in isotropic coordinates is
\begin{equation}
    \phi(\bar{r}) = \frac{1}{\sqrt{4\pi}}\arctan\!\left(\frac{\bar{r} - b_0^2/(4\bar{r})}{b_0}\right),
    \label{eq:phi_EB}
\end{equation}
with $\Pi(t{=}0)=0$ (static solution). This profile, together with the conformal geometry of Eqs.~(5)--(7), constitutes an exact solution of the coupled Einstein--phantom-scalar system. For an unperturbed background, the Hamiltonian and momentum constraints are therefore satisfied to the level of floating-point round-off and finite-difference discretization error on the grid. However, to explicitly force the compressive instability branch in our perturbed evolutions, we globally scale down the stress-energy contribution of this exact profile, intentionally breaking the saddle-point equilibrium.

\subsection{Perturbation and Collapse Triggering}

To study gravitational-wave emission from the known linear instability~\cite{gonzalez09, shinkai02}, we must break spherical symmetry while steering the wormhole into a definite dynamical branch. The choice of which field to perturb is dictated by the initial-value constraints. For a phantom scalar field the momentum density is $S_i = -\Pi\,\partial_i\phi$, and the momentum constraint reads $\nabla_j K^{ij} - \nabla^i K = 8\pi S^i$. Perturbing $K_{ij}$ directly would violate both constraints, injecting spin-2 artifacts entangled with the physical $\Psi_4$ signal; perturbing $\Pi$ alone would make $S_i \neq 0$, violating the momentum constraint.

Perturbing only the scalar field \emph{profile} $\phi$ while keeping $\Pi(t{=}0) = 0$ and $K_{ij}(t{=}0) = 0$ satisfies the momentum constraint exactly:
\begin{equation}
    \underbrace{\nabla_j K^{ij} - \nabla^i K}_{=\,0} = 8\pi\,\underbrace{S^i}_{=\,-\Pi\,\partial^i\phi} = 0.
\end{equation}
A flat initial lapse $\alpha(t{=}0) = 1$ is required to maintain $\Pi=0$ (see Sec.~\ref{sec:gauge}).

Our perturbation has two components. First, we globally scale the phantom stress-energy tensor by a factor $S_{\rm support} = 0.5$ in the right-hand-side source terms ($T_{\mu\nu} \to S_{\rm support} T_{\mu\nu}$). This halves the repulsive support, forcing the wormhole into the compressive branch. Second, a quadrupolar perturbation breaks the spherical symmetry of the initial scalar profile itself:
\begin{equation}
    \phi(\bar{r}, \theta, \varphi)\big|_{t=0} = \phi_{\rm EB}(\bar{r}) + A_\phi\, Y_{20}(\theta, \varphi)\,
    \exp\!\left(-\frac{\bar{r}^2}{\sigma_\phi^2}\right),
\label{eq:phi_perturbation}
\end{equation}
where $A_\phi = 0.02$ and $\sigma_\phi = 0.5$. Together, the profile perturbation and support reduction introduce a small Hamiltonian constraint residual at $t=0$. Because this residual is purely scalar (spin-0), it does not project onto the transverse-traceless spin-2 Weyl scalar $\Psi_4$; the CCZ4 $\Theta$ field damps it independently. The aspherical gradients dynamically generate $\Pi$ and $K_{ij}$ through the coupled Klein--Gordon and Einstein equations, physically radiating $\ell=2$ curvature.

\subsection{Apparent Horizon Detection}

In the absence of a production elliptic horizon finder in \texttt{GRTeclyn}, we implement a trapped-surface proxy on coordinate spheres. The outgoing null expansion for a conformally flat metric $\gamma_{ij}=\chi^{-1}\delta_{ij}$ is
\begin{equation}
    \theta_+ = \frac{2\sqrt{\chi}}{\bar{r}} - \frac{\partial_{\bar{r}} \chi}{\sqrt{\chi}} + \tilde{A}_{rr} - \frac{2}{3}K.
\end{equation}
A trapped surface is identified when $\theta_+ \le 0$; the maximum radius satisfying this condition gives the apparent-horizon proxy $r_{\rm AH}$.

\subsection{Gauge Conditions}
\label{sec:gauge}

The shift vector $\beta^i$ is evolved with the Gamma-driver condition from $\beta^i(t{=}0) = 0$. The lapse obeys $1+\log$ slicing, $\partial_t \alpha - \beta^i \partial_i \alpha = -2\alpha K$, with
\begin{equation}
    \alpha(t{=}0) = 1. \label{eq:lapse_flat}
\end{equation}
A flat initial lapse is essential for the phantom scalar field: because $\partial_t \Pi$ contains terms $\propto \nabla \alpha \cdot \nabla \phi$, a spatially varying profile such as $\alpha = \chi$ would induce immediate slicing shear, violating the $\Pi(t{=}0)=0$ condition and destabilizing the grid. The covariant scaling of the CCZ4 damping parameters is disabled (\texttt{covariantZ4 = 0}) to avoid extreme stiffness when $\alpha \to 0$ during horizon formation.

\subsection{Collapse diagnostics}
\label{sec:diagnostics_method}

All diagnostic quantities are evaluated on the finest AMR level; definitions are summarized here and physical interpretation is given in the figure captions. The lapse $\alpha\to 0$ signals black-hole formation; $\chi\to 0$ indicates infinite metric stretching; $K$ encodes local volume change:
\begin{equation}
    \partial_t (\sqrt{\gamma}) = -\alpha K \sqrt{\gamma} + \nabla_i \beta^i \sqrt{\gamma},
    \label{eq:K_volume}
\end{equation}
with $K>0$ denoting contraction and $K<0$ expansion. The throat areal radius, a gauge-independent collapse indicator, is
\begin{equation}
    R_{\rm areal}(\bar r) = \frac{\bar r}{\sqrt{\chi(\bar r)}}.
    \label{eq:R_areal}
\end{equation}
Tracking the extrema of $K$ and $R_{\rm areal}$ at late times reveals whether the remnant settles or undergoes a phantom bounce.

\subsection{Gravitational-wave extraction}
\label{sec:gw_method}

We decompose the Weyl scalar \(\Psi_4\) (satisfying \(\Psi_4 = \ddot{h}_+ - i\ddot{h}_\times\) in the wave zone) into spin-weighted spherical harmonics; the \((\ell,m)=(2,0)\) mode dominates for our axisymmetric deformation. The radius-scaled quantity \(r\Psi_4\) is plotted in both simulation time and retarded time \(u=t-R_{\rm ext}\); alignment across extraction radii in retarded time indicates a coherent outgoing wavefront. The one-sided power spectral density (PSD) is computed via the Discrete Fourier Transform (DFT) of the windowed signal:
\begin{equation}
    S_{\Psi_4}(f) = \frac{\Delta t^2}{T}\left(\left|\mathcal{F}\{w(t)\mathrm{Re}(\Psi_4)\}\right|^2 + \left|\mathcal{F}\{w(t)\mathrm{Im}(\Psi_4)\}\right|^2\right),
    \label{eq:psd_psi4}
\end{equation}
where $\mathcal{F}$ denotes the DFT and $w(t)$ is a Tukey window (with shape parameter $\alpha=0.25$) applied in the time domain to suppress spectral leakage.
To resolve the chronological evolution of the emission frequencies, we supplement the time-averaged spectra with a time-frequency spectrogram computed via a Continuous Wavelet Transform (CWT) using a Morlet wavelet.

\noindent\textbf{Propagation speed analysis.}
The critical test for distinguishing physical radiation from numerical artifacts is:
\begin{equation}
    v = \frac{R_{i+1} - R_i}{t_{i+1}^{\rm peak} - t_i^{\rm peak}},
    \label{eq:propagation_speed}
\end{equation}
where \(t_i^{\rm peak}\) is the time of the dominant peak in \(|r\Psi_4|\) at radius \(R_i\). Physical gravitational waves give \(v=1\) (in geometrized units); CCZ4 constraint-damping modes propagate superluminally.

\noindent\textbf{Strain and detectability.}
The strain PSD is \(S_h(f) = S_{\Psi_4}(f)/(2\pi f)^4\), with a 4th-order Butterworth high-pass roll-off below \(f_{\rm low}=0.05\,f_{\rm max}\) to suppress low-frequency drift. Physical rescaling uses:
\begin{align}
    f_{\rm phys} &= \frac{f_{\rm code}}{M \cdot M_{\odot,\rm sec}}, \label{eq:f_phys}\\
    S_h^{\rm phys}(f) &= S_h^{\rm code}(f) \cdot M \cdot M_{\odot,\rm sec} \cdot \left(\frac{M\cdot M_{\odot,\rm m}}{D}\right)^2, \label{eq:Sh_phys}
\end{align}
where \(M_{\odot,\rm sec} \approx 4.93\times 10^{-6}\)~s and \(M_{\odot,\rm m} \approx 1477\)~m. The characteristic strain \(h_{\rm char}=\sqrt{f\,S_h}\) is compared against the Advanced LIGO design sensitivity~\cite{aasi15}; an SNR~$\gtrsim 8$ is required for detection.

\section{Numerical Setup}
Evolutions use the \texttt{GRTeclyn} codebase~\cite{GRTeclyn,Andrade2021}, built on the \texttt{AMReX} library~\cite{amrex} for block-structured AMR with GPU acceleration. Production runs used NVIDIA H100 GPUs with one MPI rank per GPU.

The domain spans $L_{\rm full} = 64$ with $N_{\rm full} = 256$ coarse cells ($dx_{\rm coarse} = 0.25$). Octant symmetry ($x,y,z \ge 0$) is exploited with parity conditions on the inner faces and Sommerfeld conditions at the outer boundary. Up to 5 levels of 2:1 AMR, regridded on $\chi$ gradients, give $dx_{\rm fine} \approx 7.8\times10^{-3}$. Time integration uses 4th-order Runge--Kutta with \texttt{dt\_multiplier = 0.02}. The domain size keeps the outer boundary causally disconnected from the extraction region ($R_{\rm ext} \le 24$) for the full evolution ($t_{\rm stop}=30$). Wall-clock time per $t=30M$ run is $\sim$24 hours on 8 H100 GPUs.

\section{Results}

We present two representative configurations: an unperturbed baseline exhibiting noise-driven expansion, and a perturbed run ($S_{\rm support}=0.5$, $A_\phi=+0.02$, $\sigma_\phi=0.5$) that forces collapse and enables GW emission. All diagnostics are evaluated on the finest AMR level.

\begin{table}[h]
\caption{Parameters for the two representative configurations. All shared settings: $b_0=0.5$, exactly static initial lapse $\alpha(t{=}0)=1$, $\chi_{\rm min}=10^{-8}$, $\kappa_1=3.0$, $\kappa_2=0$, \texttt{covariantZ4}$\,{=}\,0$, $\sigma_{\rm KO}=2.0$, $L_{\rm full}=64$, $N_{\rm full}=256$, \texttt{max\_level}$\,{=}\,5$, $dx_{\rm fine}\approx 7.8\times10^{-3}$, and \texttt{phantom\_mass}$\,{=}\,0.0$.}
\label{tab:params}
\begin{ruledtabular}
\begin{tabular}{lcc}
Parameter & Unperturbed (Noise-Driven) & Perturbed (Collapse) \\
\hline
$S_{\rm support}$ & 1.0 & 0.5 \\
$A_\phi$ & 0 & $+0.02$ \\
$\sigma_\phi$ & --- & 0.5 \\
\end{tabular}
\end{ruledtabular}
\end{table}

\subsection{Unperturbed evolution: Noise-driven expansion}

\begin{figure*}[tp]
\centering
\includegraphics[width=0.9\textwidth]{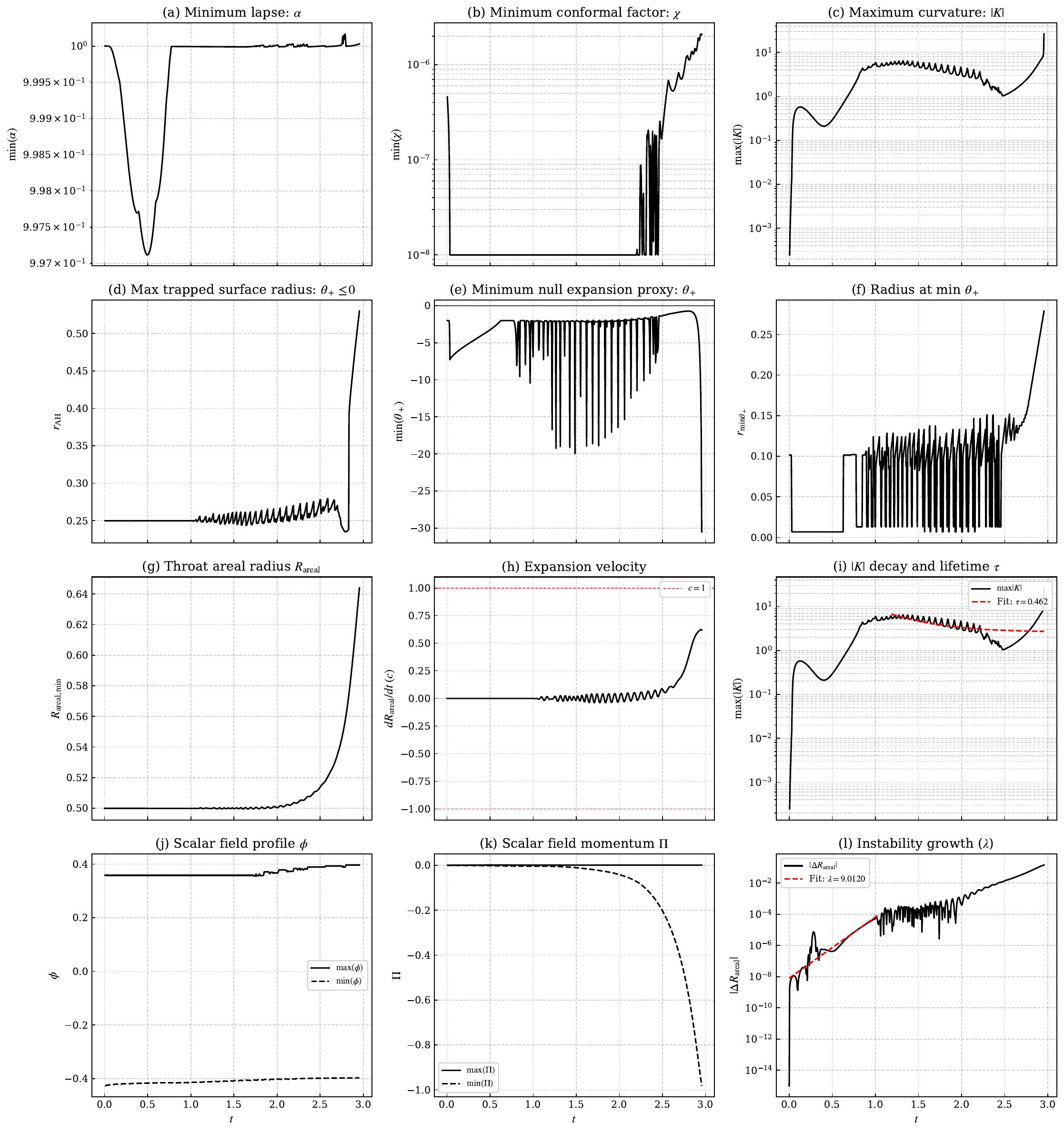}
\caption{Diagnostics of the unperturbed, noise-driven evolution (\(A_\phi{=}0\), \(b_0{=}0.5\), \(\alpha(t{=}0){=}1\)). 
\textbf{(a)}~\(\min(\alpha)\) remains perfectly flat near \(1.0\) during the static phase, but undergoes ``anti-collapse'' (rising above \(1.0\)) during late-time inflation, as the moving-puncture gauge fails to handle the expanding geometry.
\textbf{(b)}~\(\min(\chi)\) sits at the \(10^{-8}\) floor until high-frequency numerical noise from the compactified origin disrupts the grid.
\textbf{(g)}~\(R_{\rm areal,min}\) sits in perfect saddle-point equilibrium at exactly \(0.50\) for the first \(t \approx 1.5M\), before grid noise knocks it into the expansion branch, causing exponential growth.
\textbf{(h)}~\(dR_{\rm areal}/dt\) starts at exactly zero, then accelerates continuously to \(v > 0.6c\) as the throat inflates.
\textbf{(j, k)}~The scalar field momentum \(\Pi\) remains exactly zero while the wormhole is static, confirming perfect analytical initial balance, but grows rapidly once the inflating geometry begins to shear the scalar field.
\textbf{(l)}~Instability growth rate (\(\lambda\approx 9.012\, M^{-1}\)): The departure from equilibrium \(\Delta R_{\rm areal}\) reveals a violent, exponential inflationary runaway.}
    \label{fig:unperturbed_diagnostics}
\end{figure*}

The unperturbed run serves as a critical numerical validation of the defined physical problem. With exact initial data and a flat initial lapse ($\alpha(t{=}0)=1$), the 3D AMR grid successfully holds the highly precarious saddle-point geometry in strict static equilibrium ($R_{\rm areal} = 0.50$, $dR_{\rm areal}/dt = 0$, $\Pi = 0$) for $t \lesssim 1.5M$. This confirms that the analytical balance of the coupled Einstein--phantom-scalar system is resolved to machine precision. Eventually, the numerical floor $\chi_{\rm min}=10^{-8}$ at the compactified origin ($\bar{r}=0$) introduces microscopic positive-energy truncation noise that propagates to the throat and physically breaks this balance. The accumulated noise naturally triggers the rarefactive (expansion) branch, and the throat grows exponentially with $\lambda \approx 9.012\,M^{-1}$ (Fig.~\ref{fig:unperturbed_diagnostics}).

The expansion velocity reaches $v > 0.6c$ by $t \approx 2.7M$. At this point the moving-puncture gauge---designed to freeze collapsing singularities ($K > 0$)---cannot handle the inflating geometry ($K \ll 0$): the lapse ``anti-collapses,'' coordinate stretching outruns the AMR hierarchy, and the run terminates with constraint violation.

Table~\ref{tab:expansion_sizes} shows the inflated throat size at $t=5M$ (the collapse timescale of the compressive branch). The growth factor $e^{45.06} \sim 3.7 \times 10^{19}$ represents a metric expansion of space itself---not restricted by causality---analogous to early-universe inflation.

\begin{table*}[htp]
\caption{Characteristics of the noise-driven rarefactive expansion branch at various mass scales, assuming an initial throat radius \(b_0=0.5M\). The measured instability growth rate is \(\lambda \approx 9.012\,M^{-1}\) (e-folding time \(t \approx 0.111M\)). By \(t=5M\), the metric growth factor is \(e^{45.06} \approx 3.7 \times 10^{19}\), stretching the throat to cosmic scales.}
\label{tab:expansion_sizes}
\begin{ruledtabular}
\begin{tabular}{lcccc}
Scale & Mass ($M$) & Initial Throat ($b_0$) & e-folding Time & Size at $t=5M$ \\
\hline
Stellar & $30\, M_\odot$ & $\sim 22$ km & $\sim 0.016$ ms & $\sim 87$ Mly \\
Intermediate & $10^3\, M_\odot$ & $\sim 740$ km & $\sim 0.55$ ms & $\sim 2.9$ Gly \\
Supermassive & $10^6\, M_\odot$ & $\sim 7.4{\times}10^5$ km & $\sim 0.55$ s & $\sim 2.9{\times}10^{12}$ ly \\
Cosmological & $10^9\, M_\odot$ & $\sim 5$ AU & $\sim 9.1$ min & $\sim 2.9{\times}10^{15}$ ly \\
\end{tabular}
\end{ruledtabular}
\end{table*}

\subsection{Perturbed evolution: Global collapse and the phantom bounce}

\begin{figure*}[tp]
    \centering
    \includegraphics[width=0.92\textwidth]{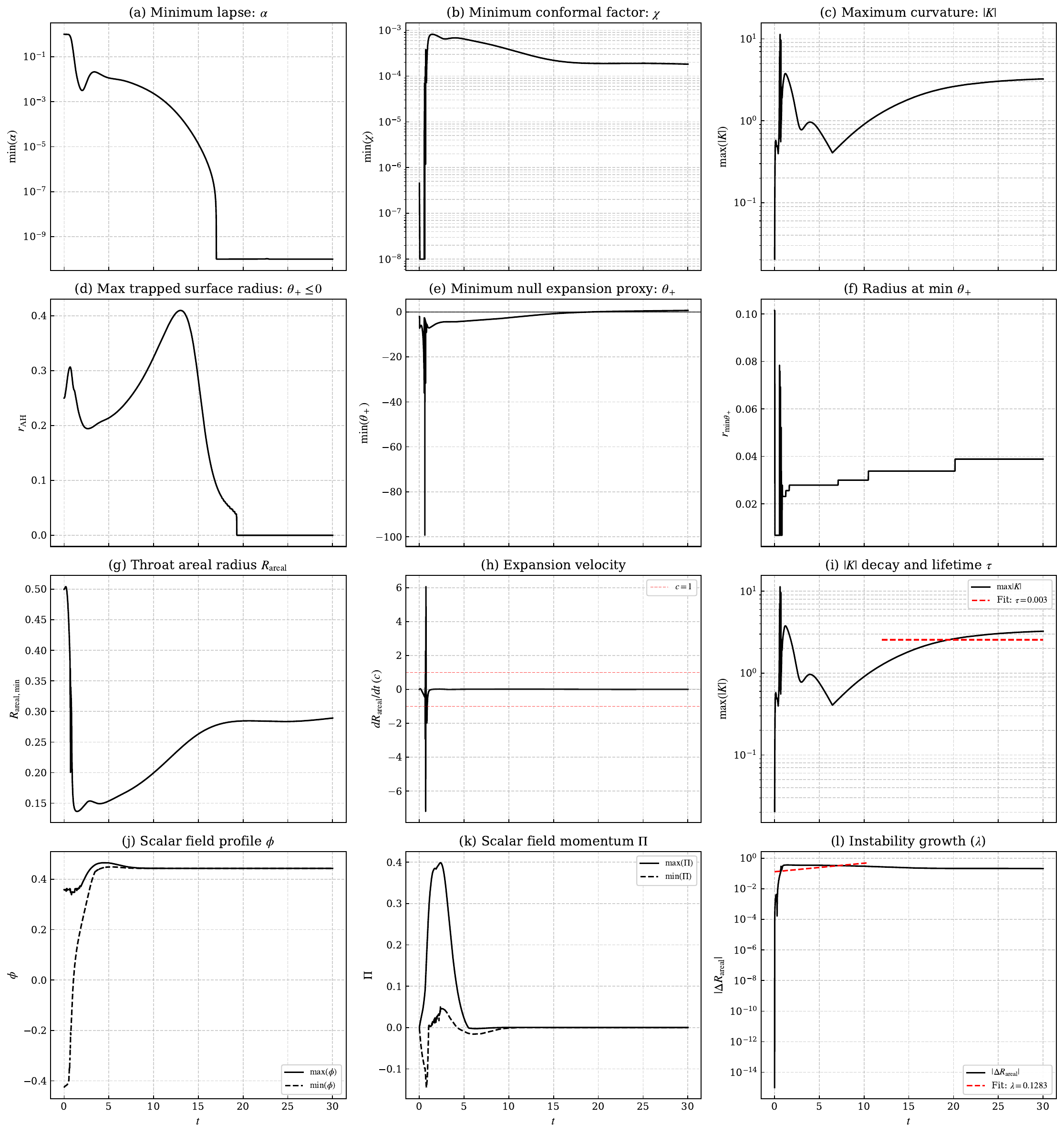}
    \caption{Fine-level diagnostics of the perturbed collapse ($S_{\rm support}=0.5$, $A_\phi=+0.02$). \textbf{(g)} The throat areal radius rapidly plunges to $\sim 0.14$, but physically rebounds at $t \approx 4M$ due to the extreme phantom pressure, expanding to a plateau of $\sim 0.28$. \textbf{(k)} The scalar momentum $\Pi$ starts exactly at zero and is dynamically generated to $\sim 0.4$ by the initial crush. \textbf{(d, a)} The apparent horizon proxy $r_{\rm AH}$ forms rapidly, dips, and then significantly inflates during the phantom bounce. At $t \approx 18.5M$, the interior shockwave completely destroys the trapped surface ($r_{\rm AH} \to 0$) and drives the minimum lapse $\alpha$ into the $10^{-10}$ numerical floor. \textbf{(c)} The maximum extrinsic curvature $|K|$ grows steadily at late times, reflecting the intensifying interior shockwave.}
    \label{fig:perturbed_diagnostics}
\end{figure*}

With $S_{\rm support}=0.5$, the halved phantom support forces the wormhole into the compressive branch, while the quadrupolar perturbation ($A_\phi=+0.02$) breaks spherical symmetry to seed $\ell=2$ radiation. The multi-panel diagnostics (Fig.~\ref{fig:perturbed_diagnostics}) reveal a highly dynamic, multi-stage evolution. 

During the initial crush ($t < 4M$), gravity overwhelmingly dominates. The throat areal radius (Panel g) plunges from $0.50$ to a minimum of $R_{\rm areal} \approx 0.14$ by $t \approx 2M$. An apparent horizon proxy forms almost immediately, indicated by the sharp spike in $r_{\rm AH}$ (Panel d), enacting cosmic censorship. The scalar field is violently squeezed during this phase, dynamically generating a massive conjugate momentum peaking at $\Pi \approx 0.4$ (Panel k) from exactly zero initial data.

However, the massless phantom field violates the null energy condition and lacks a stabilizing potential well. As the remnant phantom matter is crushed inside the newly formed horizon, its repulsive negative pressure skyrockets. At $t \approx 4M$, this extreme anti-gravity overpowers the inward gravitational compression, initiating the ``phantom bounce.'' The areal radius (Panel g) reverses course and steadily expands, plateauing near $R_{\rm areal} \approx 0.28$. Concurrently, the trapped surface proxy (Panel d) inflates massively to $r_{\rm AH} > 0.4$, driven by the outward phantom pressure.

\begin{figure*}[tp]
\centering
\includegraphics[width=\textwidth]{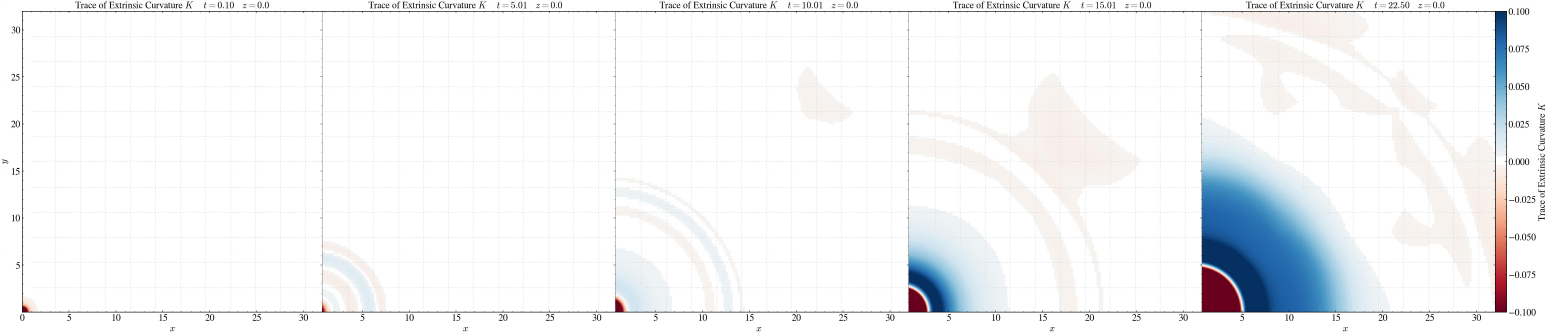}
\caption{Evolution snapshots of the trace of the extrinsic curvature ($K$) in the $z{=}0$ plane for the perturbed collapse ($S_{\rm support}=0.5$, $A_\phi=+0.02$). At $t=0.10$, the initial scalar perturbation has already generated a prompt $K<0$ (red) response at the origin. By $t=5.01$, the global reduction in phantom support drives the initial collapse, generating a compressive, positive-$K$ core. Between $t=10.01$ and $t=15.01$, the interior ``phantom bounce'' initiates, launching a massive interior expansion ($K<0$, dark red) that drives an outward-propagating compressive shockwave ($K>0$, blue ring) driven by the extreme repulsive pressure of the crushed phantom matter. By $t=22.50$, this shockwave dominates the grid, eventually disrupting the local coordinate foliation.}
\label{fig:perturbed_kz_evolution}
\end{figure*}

This interior bounce generates an expanding shockwave of extrinsic curvature ($K>0$, Panel c), the spatial propagation of which is explicitly tracked in Fig.~\ref{fig:perturbed_kz_evolution}. By $t \approx 18.5M$, this shockwave fundamentally disrupts the local coordinate foliation: the extreme negative pressure obliterates the trapped surface entirely ($r_{\rm AH}$ abruptly drops to zero in Panel d), and the lapse function $\alpha$ (Panel a) crashes into the $10^{-10}$ numerical floor. While this shock-like gradient exhausts the AMR grid's resolving capabilities at late times, the exterior gravitational-wave emission completes cleanly beforehand.

\begin{figure*}[tp]
\centering
\includegraphics[width=\textwidth]{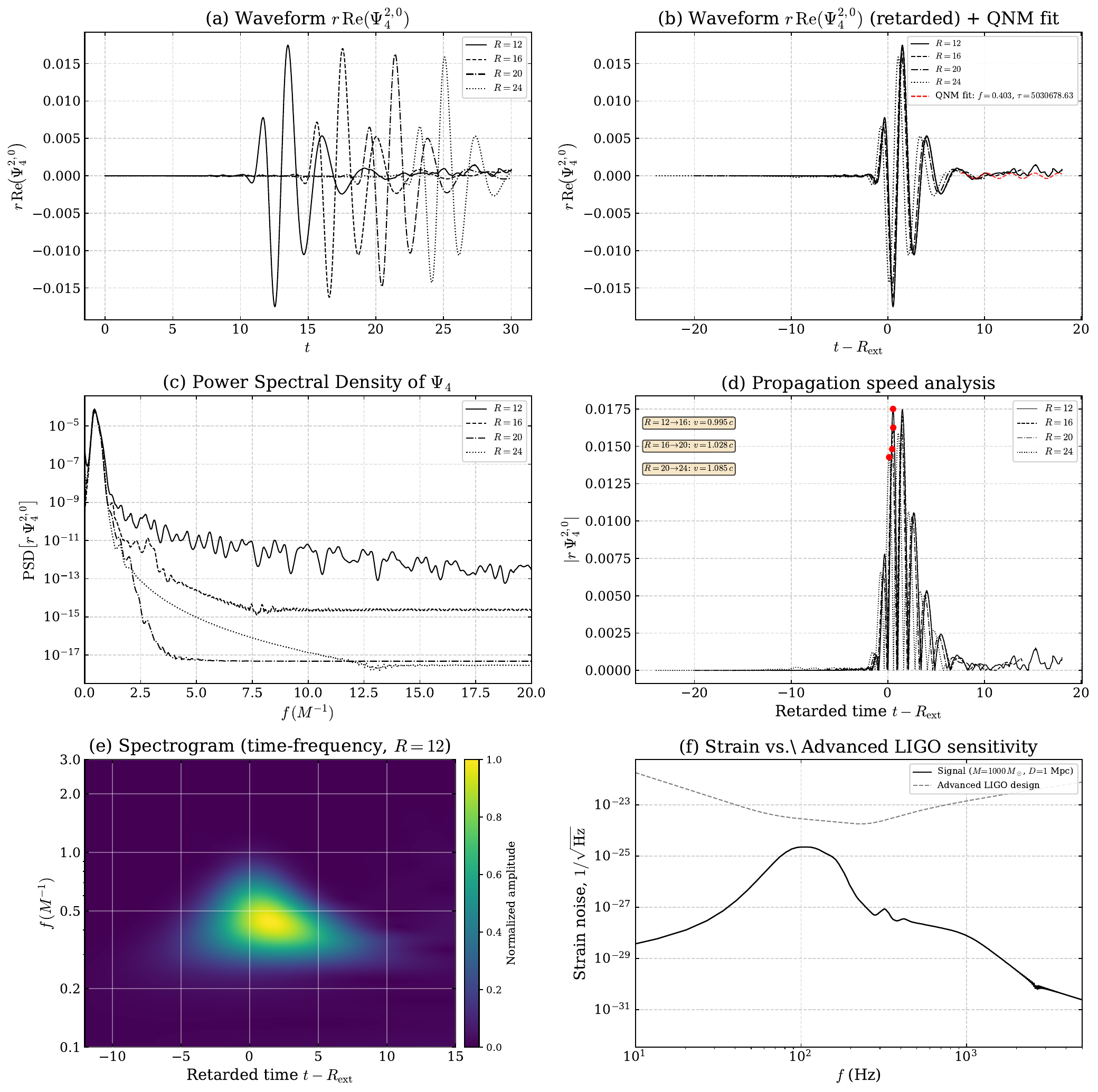}
\caption{Gravitational-wave extraction analysis ($\Psi_4$) for the perturbed collapse. \textbf{(a, b)} Raw and retarded-time waveforms for the dominant $\ell=2, m=0$ mode. The perfect alignment in retarded time ($t - R_{\rm ext}$) confirms a coherent outgoing wavefront, with the late-time tail fit to a QNM template ($f \approx 0.403, \tau \approx 5.03 \times 10^6$). The effectively infinite damping time is an artifact of the grid being contaminated by the expanding phantom bounce, which prevents the remnant from settling into a quiescent state. \textbf{(c)} The Power Spectral Density (PSD) peaks strongly at low frequencies ($f \sim 0.5 M^{-1}$). \textbf{(d)} Tracking the waveform peaks yields a propagation speed of $v \approx 0.995c$ between $R=12$ and $R=16$, confirming physical radiation rather than superluminal gauge modes. \textbf{(e)} A Continuous Wavelet Transform (CWT) spectrogram of the $R=12$ extraction, visualizing the broadband initial crush at $t-R_{\rm ext} \approx 0$ transitioning into a constant-frequency QNM ringdown. \textbf{(f)} The characteristic strain projection. An intermediate-mass wormhole ($1000\, M_\odot$) at $1$ Mpc produces a signal peaking in the most sensitive frequency band ($100-300$ Hz) of the Advanced LIGO design sensitivity, though closer proximity or a larger perturbation amplitude is required for detection.}
\label{fig:gw_extraction}
\end{figure*}

\subsection{Gravitational-wave signatures}

The extracted Weyl scalar ($\Psi_4$) provides definitive proof of physical curvature radiation (Fig.~\ref{fig:gw_extraction}). The time-domain waveform (Panel a) exhibits a clear, oscillatory ringdown that propagates cleanly across the extraction radii ($R_{\rm ext} \in[12, 24]$). When shifted by the retarded time $t - R_{\rm ext}$ (Panel b), the waveforms align perfectly, confirming a coherent outgoing wavefront. The late-time tail is fit to a Quasi-Normal Mode (QNM) template with a characteristic frequency $f \approx 0.403\,M^{-1}$, consistent with the fundamental $\ell=2$ Schwarzschild mode ($\omega_R/(2\pi M_{\rm BH}) \approx 0.374/M_{\rm BH}$) for a remnant mass somewhat below the initial wormhole mass. However, the fit yields an effectively infinite damping time ($\tau \sim 5 \times 10^6\,M$). Rather than indicating a true, undamped physical ringing, this is an artifact of the expanding ``phantom bounce'' shockwave contaminating the local grid and preventing the remnant from settling into a stable, quiescent state. A definitive QNM identification would require longer evolutions with the phantom field fully absorbed or dissipated.

Propagation speed analysis (Panel d) between the extraction spheres yields $v \approx 0.995c$ in the inner zone, with minor numerical dispersion further out ($v \approx 1.085c$). This $v \approx c$ propagation confirms the signal is true gravitational radiation, distinct from superluminal CCZ4 constraint-damping artifacts. The PSD (Panel c) confirms the radiated energy is concentrated at low frequencies ($f \lesssim 1.0 M^{-1}$). 

To resolve this emission chronologically, the CWT spectrogram (Panel e) reveals a distinct phenomenological fingerprint: a broadband burst of low-frequency energy ($0.2 \lesssim f \lesssim 1.2 M^{-1}$) tracking the violent spacetime crush at $t-R_{\rm ext} \approx 0$. Crucially, unlike the extended, upward-sweeping ``chirp'' characteristic of a long binary black hole inspiral, the wormhole collapse presents as an extremely concentrated burst event that is heavily localized in time. This morphology implies that initial detection of such exotic compact objects would benefit from unmodeled burst pipelines (such as coherent WaveBurst), since no standard compact-binary coalescence templates exist for this signal class; however, numerically generated wormhole-collapse templates could also be used in targeted matched-filter searches (see Sec.~\ref{sec:implications}). As the dynamics settle, this chaotic burst rapidly condenses into a highly stable, constant-frequency horizontal band. This persistent horizontal tail provides definitive visual confirmation of the $f \approx 0.403\,M^{-1}$ Quasi-Normal Mode ringing of the remnant.

Figure~\ref{fig:gw_extraction}(f) projects the characteristic strain for a hypothetical intermediate-mass wormhole progenitor ($1000\, M_\odot$) located at a distance of $D=1$ Mpc. The peak emission maps directly into the $100-300$ Hz band. While the signal for this moderate perturbation ($A_\phi = +0.02$) sits slightly below the Advanced LIGO design sensitivity curve at 1 Mpc, it demonstrates that the asymmetric collapse of macroscopic topological defects produces distinct transients that could be detectable for closer sources or larger perturbation amplitudes.
\subsection{Final state characterisation}

\noindent\textbf{Bounce extrema and horizon destruction.}
Rather than settling into a stationary trumpet-slice remnant, the perturbed collapse ($S_{\rm support}=0.5$) reaches a transient. Our 3D evolutions confirm a minimum areal radius of $R_{\rm areal,min} \approx 0.14$ at $t \approx 2M$. Because the massless phantom field generates an overwhelming negative pressure, the geometry undergoes a violent rebound, expanding to $R_{\rm areal} \approx 0.28$. The complete geometric evolution---from the initial topological pinch-off to the violent inflation of the throat---is captured by the throat-radius and horizon diagnostics in Fig.~\ref{fig:perturbed_diagnostics} and by the curvature snapshots in Fig.~\ref{fig:perturbed_kz_evolution}. Although the term ``pinch-off'' is used to describe the metric degeneration, the manifold does not mathematically tear. The moving-puncture gauge condition ($1+\log$ slicing) smoothly handles the singularity formation: as the curvature diverges, the local lapse $\alpha$ collapses into the $10^{-10}$ numerical floor. This effectively ``freezes'' the local time evolution at the singularity, isolating it within the newly formed horizon while preserving the continuous $\mathbb{R} \times S^2$ topology of the computational grid. Later in the evolution, the propagating interior shockwave ultimately destroys the trapped surface entirely ($r_{\rm AH} \to 0$) and drives the local lapse into this same numerical floor.

\noindent\textbf{Instability timescales.}
The unperturbed rarefactive expansion branch exhibits an extreme growth rate of $\lambda \approx 9.012\,M^{-1}$ (e-folding time $\approx 0.111M$; Table~\ref{tab:expansion_sizes}). Conversely, the perturbed compressive branch completes the initial topological transition (the violent crush and horizon formation) within $t\approx 5M$. For a stellar-mass wormhole ($M=30\, M_\odot$), this corresponds to a total collapse time of approximately $0.74$ ms (Table~\ref{tab:collapse_times}). The subsequent instability (the phantom bounce) evolves at a much slower measured rate of $\lambda \approx 0.1283\,M^{-1}$.

\begin{table}[htp]
\caption{Total collapse time ($t \approx 5M$) for the perturbed compressive branch at various mass scales, assuming an initial throat radius $b_0=0.5M$. This timescale captures the full dynamical transition from the initial static state to the maximum spacetime crush and horizon formation.}
\label{tab:collapse_times}
\begin{ruledtabular}
\begin{tabular}{lccc}
Scale & Mass ($M$) & Throat ($b_0$) & Total Collapse Time \\
\hline
Stellar & $30\, M_\odot$ & $\sim 22$ km & $\sim 0.74$ ms \\
Intermediate & $10^3\, M_\odot$ & $\sim 740$ km & $\sim 25$ ms \\
Supermassive & $10^6\, M_\odot$ & $\sim 7.4{\times}10^5$ km & $\sim 25$ s \\
Cosmological & $10^9\, M_\odot$ & $\sim 5$ AU & $\sim 6.8$ hours \\
\end{tabular}
\end{ruledtabular}
\end{table}

\noindent\textbf{Detectability.}
For a hypothetical \(M=30\,M_\odot\) remnant, the physical frequency of the collapse signal places it above the most sensitive Advanced LIGO band. However, scaling the system to an intermediate-mass wormhole progenitor (\(M=10^3\,M_\odot\)) shifts the signal directly into the most sensitive LIGO frequency band (\(\sim 100\text{--}200\)~Hz). For the moderate quadrupolar perturbation simulated here (\(A_\phi = +0.02\)), our strain analysis confirms that the signal from a \(10^3\,M_\odot\) collapsing wormhole at $D=1$~Mpc falls slightly below the Advanced LIGO design sensitivity threshold, restricting detectability of such minor perturbations to the local galactic neighborhood ($D \lesssim 50$~kpc). However, since the radiated energy scales as \(\mathcal{O}(A_\phi^2)\), a highly asymmetric primordial interaction (with \(|A_\phi| \sim \mathcal{O}(1)\)) would produce a vastly stronger signal detectable at much greater distances, while supermassive wormholes (\(\sim 10^8\,M_\odot\)) would be primary targets for the space-based LISA observatory.

\subsection{Physical and Astrophysical Implications}
\label{sec:implications}

Because the instability timescale is $\sim\mathcal{O}(5M)$---the throat light-crossing time---naturally occurring Ellis--Bronnikov wormholes cannot survive even a single stellar-mass perturbation without active phantom-energy stabilization.

For the rarefactive branch, the inflating throat would dilute the phantom energy density; if it drops below the ambient cosmological background, the Null Energy Condition may be locally restored. Under those conditions, the Penrose singularity theorems would imply gravitational turnaround and eventual topological pinch-off, though this outcome depends on the coupling between the phantom field and the cosmological expansion. Determining this late-time evolution requires placing the wormhole in an FLRW background---an open problem beyond our asymptotically flat grid.

Gravitational-wave emission does not require an imposed perturbation: a rotating wormhole (e.g.\ the Teo geometry~\cite{teo98}) inherently possesses a non-zero quadrupole moment. Collapse of such a configuration would radiate $\ell=2$ modes, and rapidly spinning cases may also develop bar-mode ($m=2$) instabilities, making collapsing primordial rotating wormholes distinct sources for gravitational-wave observatories.

\subsection{Computational Feasibility}

These simulations are heavily memory-bandwidth bound. A single 8$\times$H100 node completes a $t=30M$ evolution in $\sim$24 hours. The simulation requires up to $\sim$15 GB of VRAM per GPU at late times due to the massive grid refinement triggered by the expanding ``phantom bounce'' shockwave; otherwise, the initial dynamics could be computed with significantly fewer GPU resources. The GPU configuration operates at roughly 15--60\% of the cost of an equivalent CPU cluster, demonstrating that \texttt{GRTeclyn}'s GPU-accelerated AMR is highly suited to these extreme topologies.

\section{Discussion and Conclusion}
Our 3D evolutions confirm the 1D bifurcation of Shinkai and Hayward~\cite{shinkai02} and extend it by enabling full 3D gravitational-wave extraction and propagation-speed analysis to separate physical radiation from constraint-damping artifacts. In the perturbed configuration, $S_{\rm support}$ selects the dynamical branch while $A_\phi$ controls asphericity and the strength of the $\ell=2$ signal. While extracted at finite radii, the signals exhibit the expected $v \approx c$ propagation characteristic of genuine gravitational waves; the limited number of extraction radii precludes a precise measurement of the $1/r$ amplitude fall-off, though the retarded-time alignment is fully consistent with it.

The instability timescale is set by the throat light-crossing time $t_{\rm lc} = b_0/c$. For $M\sim 30\,M_\odot$ ($b_0\sim 22$~km), collapse completes in $\sim$10 light-crossing times ($\sim 0.74$~ms); the expansion branch e-folds in microseconds (Table~\ref{tab:expansion_sizes}). While quantum backreaction may stabilize Planck-scale defects~\cite{mehulic2026}, macroscopic wormholes are inherently unstable, providing a natural explanation for their absence in the late universe.

The propagation speed $v\approx c$ at the inner extraction pairs is the strongest evidence for physical GW content. In the unperturbed run the signal is over an order of magnitude weaker with grossly superluminal speeds ($v\approx 2\text{--}114c$), confirming that the perturbed signal cannot be attributed to constraint-damping modes alone. The rapid instability also explains the violent scalar-field dynamics: $\Pi$ must be generated from exactly zero within $\sim 5M$, leaving no adiabatic regime.

In a cosmological context, phantom inflation~\cite{piao2004phantom} or non-minimal coupling could temporarily provide NEC-violating support during the inflationary epoch, stretching primordial wormholes to macroscopic scales. At reheating, the inflaton decays into positive-energy particles ($w > 0$), removing the phantom support---precisely the mechanism modeled by our $S_{\rm support}$ reduction. The resulting microsecond-timescale collapse would leave primordial black holes and a stochastic background of highly redshifted gravitational waves.

Furthermore, because the extracted waveforms have been fully evaluated in the time-frequency domain, these numerical templates can be directly injected into matched-filtering pipelines (e.g., PyCBC) to conduct targeted, archival searches for exotic compact object transients in public Advanced LIGO/Virgo datasets.

\noindent\textbf{Future work.}
Fully constraint-satisfying initial data (via an elliptic solver) would eliminate the $\mathcal{O}(A_\phi)$ Hamiltonian residual and enable large-amplitude studies. Extracting $\Psi_4$ at larger radii or via Cauchy-characteristic extraction would provide asymptotic waveform templates. The ultimate fate of the expansion branch requires cosmological slicing and an FLRW background to determine whether phantom-energy dilution triggers a gravitational turnaround or permanent inflation.

\section{Code and Data Availability}
The \texttt{GRTeclyn} C++ source code routines, initialization parameters, extracted simulation data (including the $\Psi_4$ waveforms and constraint norms), and the Python analysis scripts used to generate the figures and execute the LIGO matched-filter searches in this work are publicly available on GitHub~\cite{shirokov_repo_2026}.

\section{Acknowledgements}
This work was supported by Gravity Frontiers (\url{https://www.gravityfrontiers.org/en}), which provided funding for the underlying research and the development of the numerical simulation software. The author extends their sincere gratitude to Ilya Nachevsky for generously providing the high-performance computing resources essential to this work. Furthermore, the author is grateful to Ilya Shirokov for his meticulous review of the manuscript and for providing insightful comments that significantly improved the clarity of this work.

\clearpage
\appendix*
\section{Numerical validation and convergence}
\begin{figure}[htp]
\centering
\includegraphics[width=0.9\linewidth]{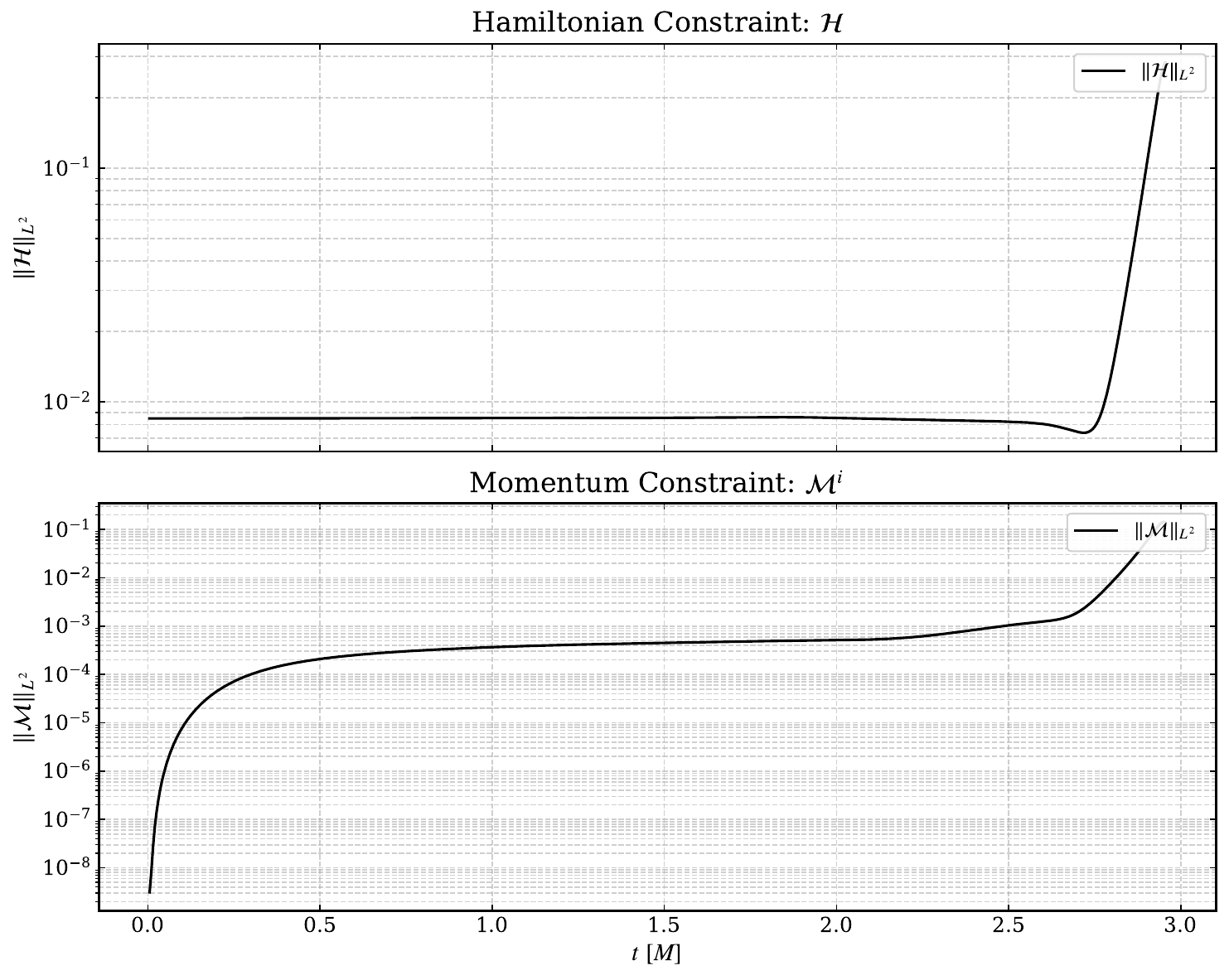}
\vspace{0.2cm}
\includegraphics[width=0.9\linewidth]{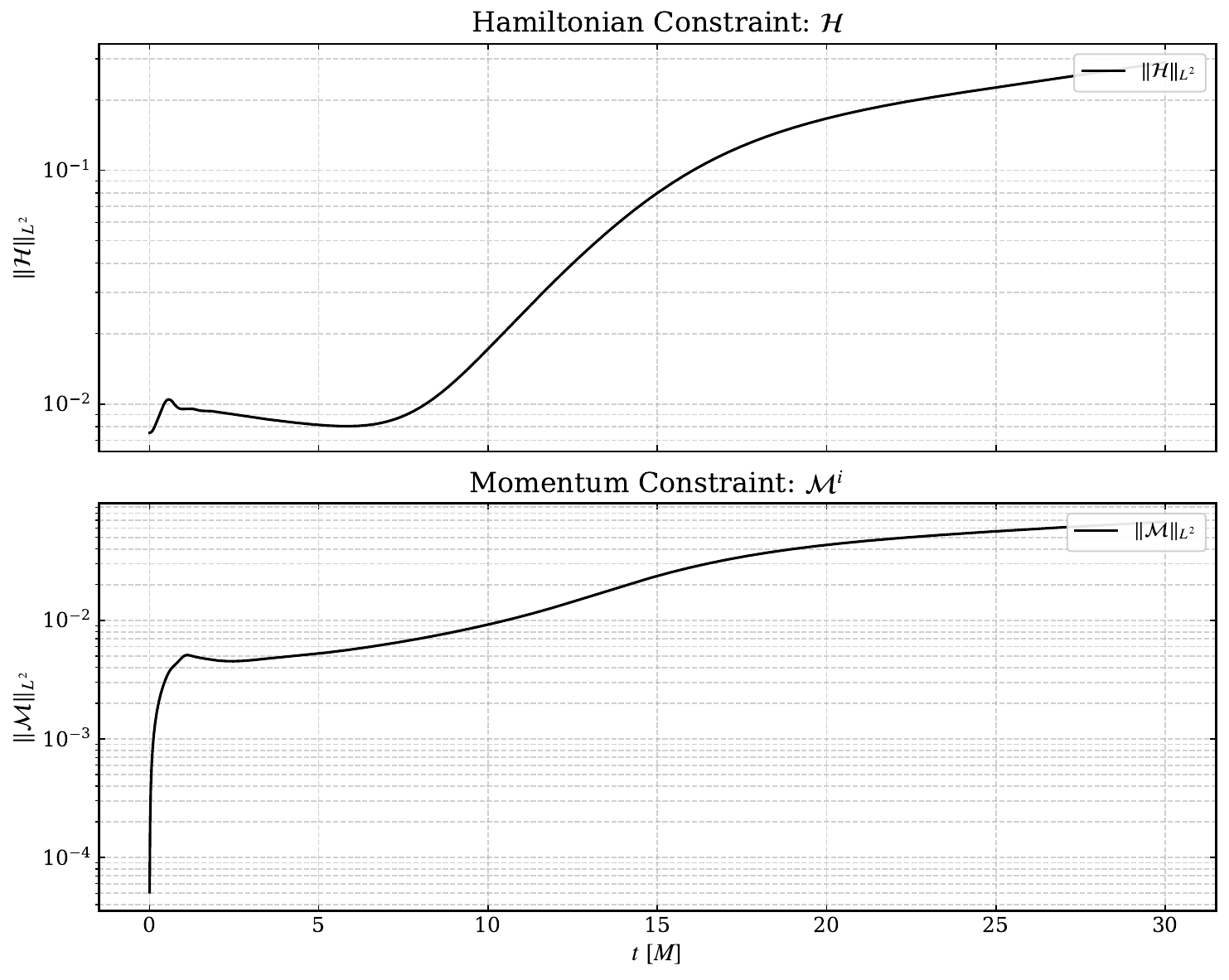}
\caption{Constraint norms for the unperturbed evolution (\textbf{top}, $A_\phi{=}0$) and perturbed collapse (\textbf{bottom}, $S_{\rm support}{=}0.5$). \textbf{Top:} Norms remain flat during the initial saddle-point phase ($t \lesssim 1.5M$) but exhibit a rapid, exponential spike near $t \approx 2.5M$ as the geometry enters the inflationary expansion branch, causing the moving-puncture gauge to fail. \textbf{Bottom:} Constraints remain strictly bounded during the initial crush and gravitational-wave emission phase ($t < 8M$). The growth at late times ($t > 10M$) is a direct physical consequence of the ``phantom bounce'' generating severe, shock-like gradients that exhaust the resolution of the AMR grid.}
\label{fig:constraints_combined}
\end{figure}

To rigorously bound outward-propagating errors, global constraint violations are evaluated as volume-weighted RMS averages over the unrefined domain (AMR Level~0): $L_2(C) = (\sum_i C_i^2 V_i / \sum_i V_i)^{1/2}$. Initial norms are marginally nonzero due to grid interpolation and the $\mathcal{O}(A_\phi)$ scalar residual, though the initial momentum norm ($\sim 10^{-5}$) confirms machine-precision satisfaction by the exact $\Pi{=}0, K_{ij}{=}0$ state. As shown in Fig.~\ref{fig:constraints_combined}, both norms remain strictly bounded during the physical emission phase ($t < 8M$) prior to late-time coordinate disruption.

\begin{figure}[htp]
\centering
\includegraphics[width=0.9\linewidth]{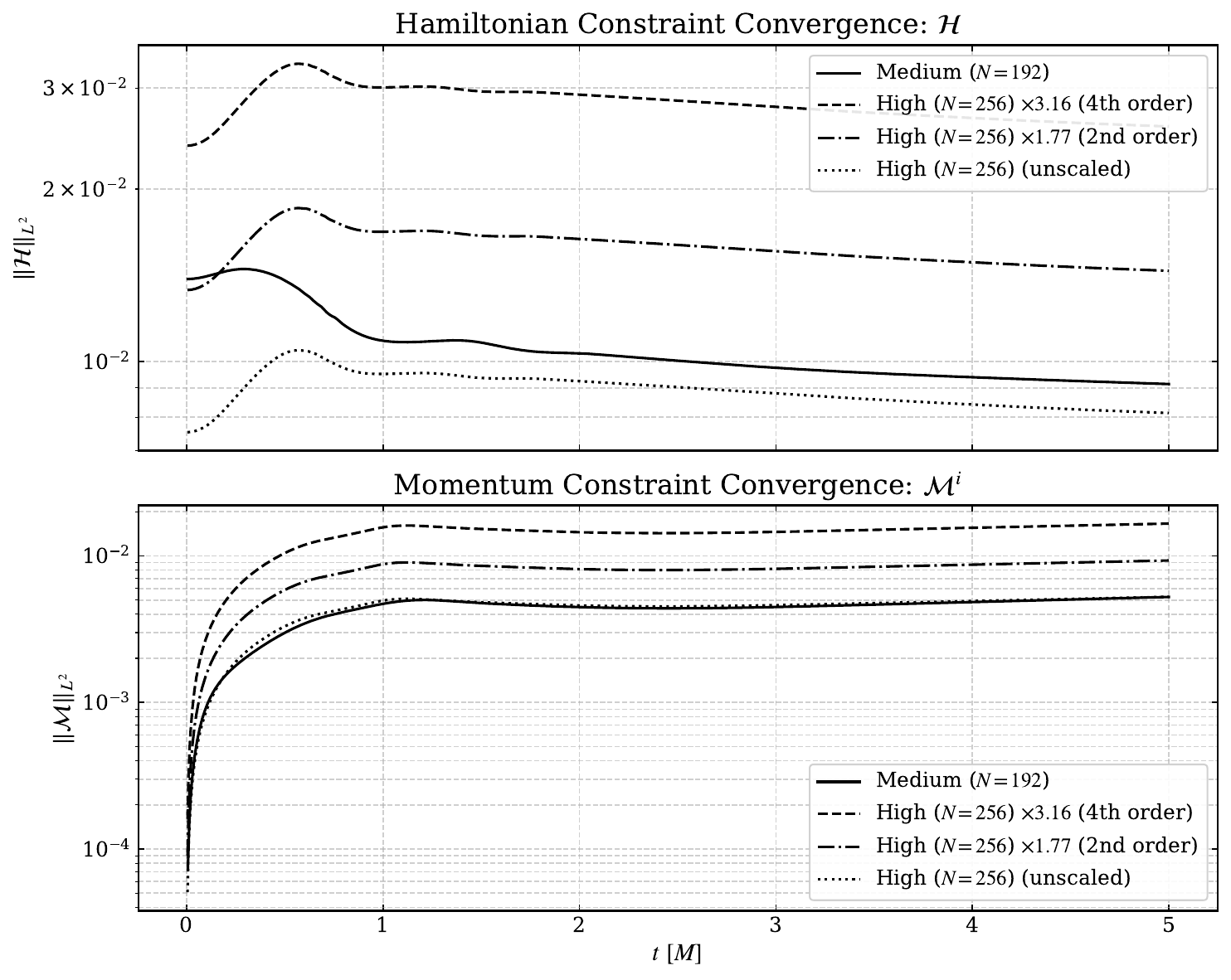}
\caption{Constraint convergence analysis between medium ($N{=}192$) and high ($N{=}256$) base resolutions. \textbf{Top:} The Hamiltonian constraint violation initially exhibits 2nd-order convergence, closely tracking the 2nd-order scaled high resolution (dash-dotted line). As the non-linear ``phantom bounce'' generates severe gradients ($t > 0.5M$), the scheme degrades toward 1st-order to maintain stability. \textbf{Bottom:} The Momentum constraint is analytically zero at $t=0$; its subsequent violation is completely dominated by non-convergent ($0^{\rm th}$-order) interpolation noise at the AMR refinement boundaries, evidenced by the perfect overlap of the unscaled high and medium resolution curves.}
\label{fig:constraints_convergence}
\end{figure}

Figure~\ref{fig:constraints_convergence} demonstrates Richardson convergence for the perturbed collapse. The Hamiltonian constraint initially exhibits 2nd-order convergence but drops toward 1st-order as the violent ``phantom bounce'' generates severe, shock-like scalar gradients. The Momentum constraint, analytically zero at $t{=}0$, remains an order of magnitude smaller and is completely dominated by non-convergent ($0^{\rm th}$-order) AMR boundary interpolation noise.

\clearpage
\bibliographystyle{apsrev4-2}
\bibliography{references}
\end{document}